\documentclass{ws-procs9x6}

\def\gsimeq{\mathrel{\hbox{\rlap{\hbox{\lower4pt\hbox{$\sim$}}}\hbox{$>$}}}}
\def\lsimeq{\mathrel{\hbox{\rlap{\hbox{\lower4pt\hbox{$\sim$}}}\hbox{$<$}}}}

\def\xmm{{\tt XMM}--{\it Newton}}
\def\chandra{{\it Chandra}}

\def\beppo{{\it BeppoSAX}}

\newcommand{\apj}[2]{\mbox{ {\em Ap.J.\ }{\bf #1}, {#2}}}
\newcommand{\aj}[2]{\mbox{ {\em A.J.\ }{\bf #1}, {#2}}}
\newcommand{\apjs}[2]{\mbox{ {\em Ap.J.S.S.\ }{\bf #1}, {#2}}}
\newcommand{\aap}[2]{\mbox{ {\em A.\&A.\ }{\bf #1}, {#2}}}
\newcommand{\mnras}[2]{\mbox{ {\em M.N.R.A.S.\ }{\bf #1}, {#2}}}

\begin{document}

\title{The missing X-ray background}

\author{ANDREA COMASTRI}

\address{INAF--Osservatorio Astronomico di Bologna, \\
via Ranzani 1, I--40127 Bologna, Italy \\
E-mail: andrea.comastri@bo.astro.it}


\maketitle

\abstracts{The fraction of the hard X-ray background ({\tt XRB}) 
resolved into individual sources by the deep \chandra\  and \xmm\  
surveys strongly depends on the adopted energy range and decreases 
with increasing energy. As a consequence the nature of the 
sources of the even harder ($>$ 10 keV) {\tt XRB} remains observationally
poorly constrained. I will briefly discuss the need for  
X--ray observations above 10 keV.}

\section{Introduction}

After the impressive achievements obtained by  \chandra\ and \xmm\ 
in terms of angular resolution and high energy throughput, almost all the 
papers dealing with X--ray observations of extragalactic sources 
begin with the following general statement: {\it ... Deep  \chandra\ 
and \xmm\ surveys have resolved most of the {\tt XRB} into discrete
sources...} While there are no doubts about the origin 
of the {\tt XRB}, the above statement is true only when referred to 
the 2--10 keV energy range and it is even more true around 2--3  keV
rather than above 7--8 keV, where the resolved fraction is no more
than 50\% \cite{worsley}.
At energies greater than 10 keV, where the bulk of the {\tt XRB} 
energy density is produced, the resolved fraction is negligible, 
being strongly limited by the lack of imaging X--ray observations at high
energies. As far as the 10--100 keV band is concerned, we are currently 
facing the problem already encountered for the 2--10 keV band 
right after a significant fraction of the 1 keV background 
was resolved by {\it Einstein} \cite{riccardo} surveys.

The most important difference is that a solid model for the {\tt XRB},
based on the AGN unification scheme, is now available.
First proposed by Setti and Woltjer \cite{setti} and 
elaborated with an increasing level of details since that time
\cite{madau} \cite{comastri} \cite{gilli} the {\tt XRB} synthesis 
is obtained assuming a dominant contribution of obscured AGN with 
a wide range of column densities and luminosities. 
Though differing in several details regarding the luminosity 
function and X--ray spectral shape parameterization, the different 
flavours of synthesis models, published so far, share the same key feature:
the X--ray obscuration. In the following I will refer to these 
as {\tt absorption models}.

 \begin{figure*}[t]
       \psfig{file=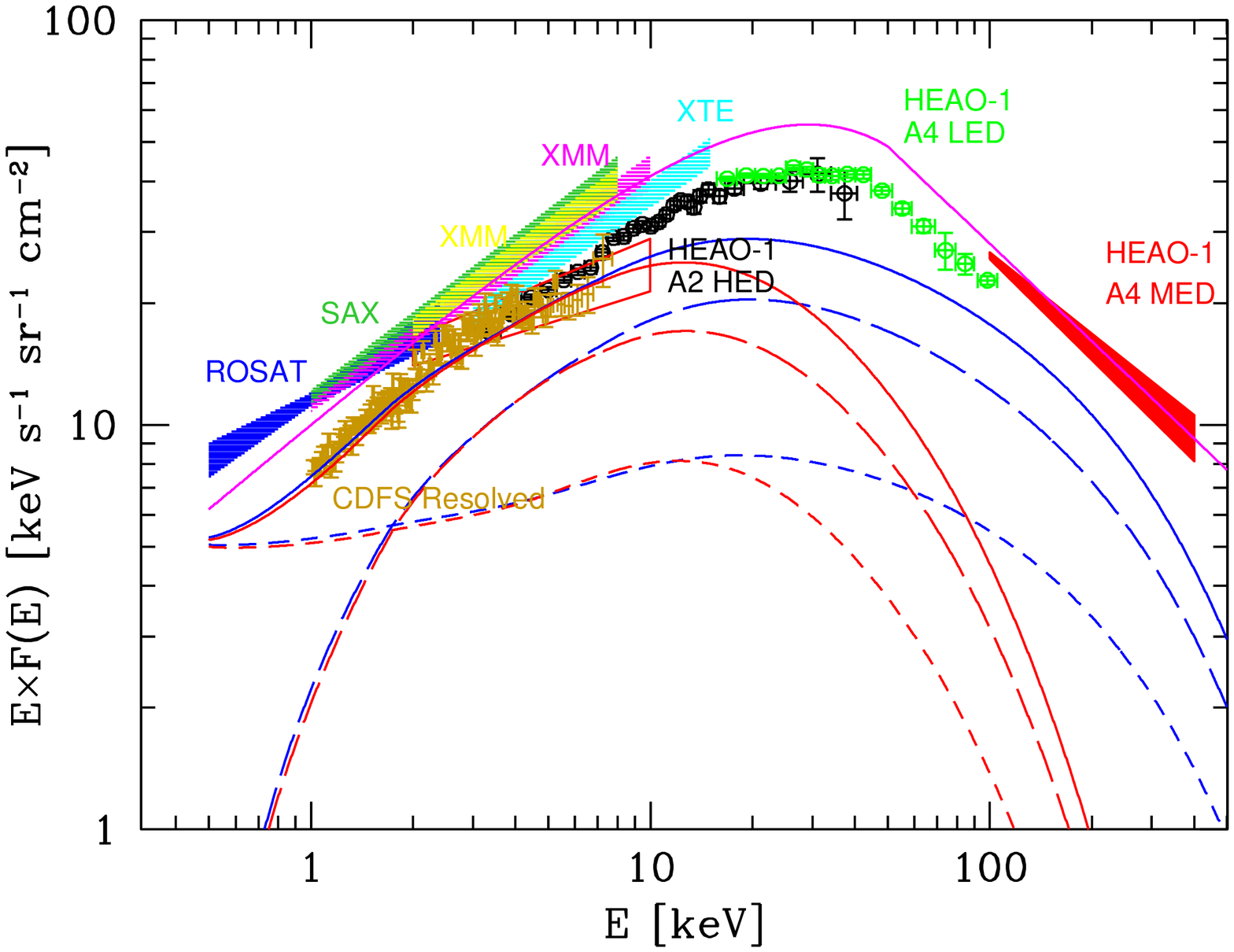,width=4.7in}
\caption[]{A selection of XRB spectral measurements collected from 
observations with different satellites as labeled. ROSAT blue \cite{ioannis};
 \beppo\ green \cite{vecchi}; \xmm\ 
magenta \cite{lumb} and yellow \cite{deluca}; RXTE 
cyan \cite{revni}; HEAO1--A2 black \cite{gruber} and HEAO1--A4 LED green 
\cite{gruber}; HEAO1--A4 MED red \cite{kinzer}.
Also reported is the integrated contribution 
of resolved sources in the \chandra\ CDFS \cite{tozzi} 
(gold points)
and in \xmm\ Lockman hole \cite{worsley} (red bow--tie).   
The solid magenta curve represents the analytical fit of Gruber 
\cite{gruber} 
renormalized upward by 30\% in order to fit the most recent measurements.
The blue ($E_{cut}$ = 400 keV) and red ($E_{cut}$ = 100 keV) 
solid curves represent the integrated contribution of the model 
described in the text. The short--dashed curves correspond to unabsorbed 
AGN ($logN_H < 22$ cm$^{-2}$), while the long--dashed curves 
correspond 
to obscured Compton--thin sources ($22 < logN_H < 24$ cm$^{-2}$).}
   \label{}
   \end{figure*}

Massive campaigns of multiwavelength follow--up  observations 
have made possible to obtain spectroscopic and photometric redshifts
for several hundreds of hard X--ray (2--10 keV) 
selected \chandra\ and \xmm\ sources \cite{barger} \cite{szokoly}
(see Tables 1 and 2 in Brandt et al. \cite{brandt} for a summary). 
The discovery of a sizable fraction of X--ray obscured
sources agrees, at least to a first approximation, with the 
{\tt absorption models} predictions. However the observed 
absorption and redshift distributions are poorly reproduced 
by current models.

Although a rather obvious way to cope with these problems 
is to construct {\tt absorption models} with different luminosity and
absorption functions until a better agreement with the overall
observational constraints is achieved (see for instance Ueda et al. 
\cite{ueda}), 
a relatively large region of the
parameter space remains unaccessible due to the lack of
observational constraints.

Within the framework of {\tt absorption models}, 
the shape of the {\tt XRB} spectrum and intensity 
in the 10--100 keV range, where most of its energy density is 
contained, is modeled assuming an important contribution 
from heavily obscured Compton thick ($N_H > 1.5 \times 10^{24}$ cm$^{-2}$) 
sources around 20--30 keV. Moreover a high energy 
cut--off ($E_{cut}$) 
usually parameterized as an exponential roll--over with an 
e--folding energy of the order of a few hundreds of keV has to be present 
in the high energy spectrum of all the sources in order 
not to overproduce the observed {\tt XRB} above 100 keV.

The most reliable observational constraints on the
Compton thick AGN space density and the exponential
cut--off energy have been obtained thanks to the 
{\tt PDS} instrument on board \beppo\ . Though limited to 
bright 10--100 keV fluxes, the \beppo\  results indicate that
a fraction as high as 50\% of the
Seyfert 2 galaxies in the nearby Universe are obscured by Compton
thick gas \cite{risaliti}, while the e--folding energy  
in the exponential cut--off spans a range from about 80 to 
more than 300 keV \cite{perola}.

In the following I will briefly discuss the impact that 
different assumptions on the fraction of 
Compton thick sources and $E_{cut}$ values of the continuum have 
on the synthesis of the high energy {\tt XRB} spectrum.  
Given that the emission processes reponsible of Compton thick 
absorption and high energy cut--off 
are basically driven by Compton scattering, I will refer to 
this approach as {\tt Compton models}.

\section{Resolved and Unresoved XRB}

For the purposes of the present exercise I will adopt the
parameters used by Comastri et al. (1995).
Although there are compelling evidences which indicate that
the evolution of the X--ray luminosity function could not be 
parameterized by pure luminosity evolution anymore
simple as postulated in the simplest versions of 
{\tt absorption models}, \cite{ueda}  it is important
to point out that the most recent findings are not expected to substantially
modify the model predictions above 10 keV.

  \begin{figure*}[t]
       \psfig{file=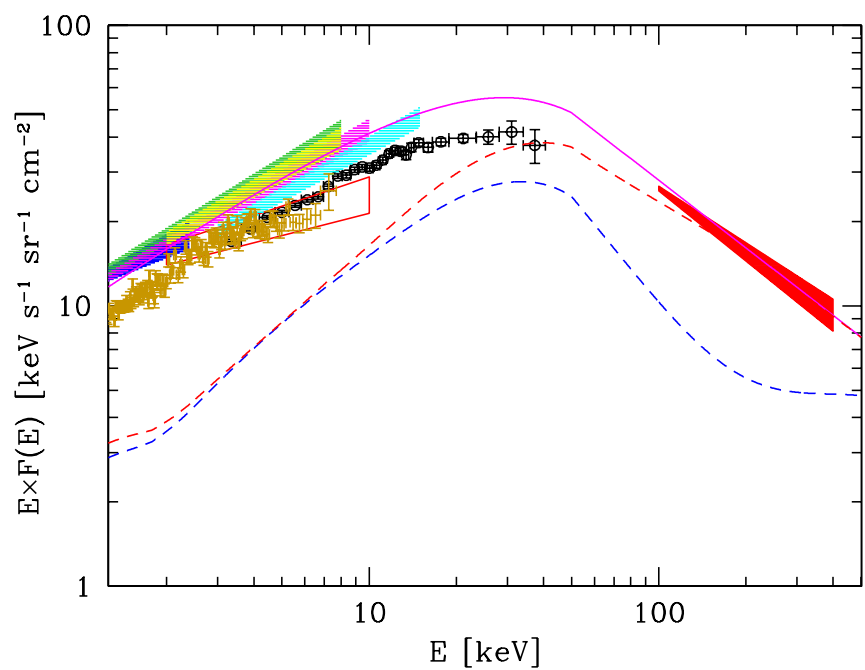,width=4.7in}
\caption{The residual background spectrum 
after subtraction from the two models reported 
in Fig.~1 
(red line for $E_{cut}$ = 100 keV, blue line for $E_{cut}$ = 400 keV).
Data as in Figure 1. The HEAO1--A4 LED dataset (green points in 
Fig.~1) are not reported for clarity.}
   \label{}
   \end{figure*}

The starting and admittely extreme assumption  
is that Compton thick absorption is not energetically relevant.
Since the observational constraints on the presence of 
Compton thick sources beyond the local Universe are rather poor
and only a few examples have been reported (see Comastri \cite{comastri1} 
for 
a review), hereafter I will consider only the effects of 
two representative values (100 and 400 keV)  
for the high energy cut--off.
It is possible to account for the resolved fraction of the 
1--10 keV background as measured
by \chandra\ \cite{tozzi} and \xmm\ \cite{worsley}
with an appropriate, but reasonable, tuning of the absorption distribution.
More specifically the ratio between Compton thin and unabsorbed 
AGN (assuming log$N_H <$ 22 as the dividing line) 
is in the range 2.0--2.4 (for $E_{cut}$=400 and 
100 keV respectively) to be compared with 2.8 in the {\tt absorption model} 
of Comastri et al. (1995). 
The resulting model spectra are reported in Fig.~1 along with 
a compilation of {\tt XRB} data which clearly demonstrates a mismatch 
between the first HEAO1--A2 measurement and the recent estimates in the 
2--10 keV band. The maximum deviation is of the order of 40\% at 10 keV.
For the purposes of the present discussion I consider the analytical 
fit of Gruber (1992) renormalized upward by a factor 1.3 and slightly 
modified above about 60 keV to match the HEAO1--A4 MED spectrum. 
This approximation 
results in a quite good description of the 2--10 keV \xmm\ data and 
settles in between \beppo\ and {\it Rossi}--{\tt XTE} observations.
Though the renormalized Gruber analytical fit provides a reasonable 
description of the broad band {\tt XRB} spectrum a few words of caution 
are appropriate. A closer look at the {\tt XRB} data points 
above the peak indicates, besides cross--calibration uncertainties 
between the A4 LED/MED experiments, a sharp break at $E >$ 40 keV which
is responsible for the knee in the analytical approximation. Moreover 
while it is obvious that the peak in the {\tt XRB} spectrum must be around 
a few tens of keV, the exact location and intensity may still be subject
to significant uncertainties, especially after the recent measurements
below 10 keV which cast doubts on the normalization of the HEAO1--A2 
data. The residual spectra (Fig.~2), 
computed subtracting the model predictions 
from the renormalized ``observed'' {\tt XRB} intensity, are subjected to the
caveats illustrated above. 

\section{Conclusions}

Not surprisingly, the shape of the model dependent residual background is 
reminiscent of that of extremely obscured sources. 
While the contribution of a yet uncovered population of high redshift
Compton thick sources \cite{fabian} 
may contribute to fill the gap, it is interesting 
to note that the residual spectrum peaks at energies of the order of
30--50 keV (depending upon the assumed $E_{cut}$).  
If the peak is associated to a characteristic energy in 
the source spectra, then for a typical redshift of $\simeq$ 1 
this would correspond to 60--100 keV rest--frame. These energies 
cannot be obtained increasing absorption because at high column densities 
Compton down--scattering strongly depresses the entire high energy spectrum.
An alternative possibility implies high energy cut--off values 
clustered within a relatively narrow range which in turns depends 
from the average redshift ($E_{cut} \sim$ 50--100 keV for 
$\langle z \rangle \simeq 1$) of the sources. 
Such a population has to be relatively
well localized in space in order to reproduce the rather sharp 
spectral break of the residual spectrum (especially pronounced 
if $E_{cut}$ = 400 keV, blue curve in Fig.~2) which otherwise 
would be smeared out integrating over the cosmological volume. 

Taking at the face value the results above described
it is also possible that the sources of the residual background
are characterized by a truly flat spectrum ($\Gamma \simeq$ 1.1--1.2) 
up to several tens of keV 
which breaks to a much steeper ($\Gamma \simeq$ 2.6--2.9) 
slope (not necessary an exponential cut--off) at higher energies. 
Although it seems premature to invoke 
a new population of sources with a peculiar hard X--ray spectrum the 
present exercise highlights the need for X--ray observations in the
$E >$ 10 keV domain.

\section*{Acknowledgments}
I kindly acknowledge support by INAOE, Mexico, during the
2003 Guillermo--Haro Workshop where part of this work was performed.
Partial support from ASI I/R/057/02 and MIUR Cofin--03--02--23 grants 
is also acknowledged. Giancarlo Setti, Gianni Zamorani and Cristian 
Vignali are acknowledged for a careful reading of the manuscripts and
illuminating comments.
A special thank to Raul P.M. Mujica and Roberto P. Maiolino for organizing
a stimulating and exciting workshop and for their patience.

\end{document}